\documentclass[prl,twocolumn]{revtex4}
\usepackage{epsfig}
\begin{document}
\title{Fermion Superfluids of Non-Zero Orbital Angular Momentum near Resonance}
\author{Tin-Lun Ho and Roberto B. Diener}
\address{Department of Physics,  The Ohio State University,
Columbus, Ohio 43210}

\begin{abstract}
We study the pairing  of Fermi gases near the scattering resonance of the $\ell\neq 0$ partial wave. Using a model potential which reproduces the actual two-body low energy scattering amplitude, we have obtained an analytic solution of the gap equation. We show that the 
ground state of $\ell=1$ and $\ell=3$ superfluid are orbital ferromagnets with pairing wavefunctions
 $Y_{11}$ and $Y_{32}$ respectively. For $\ell=2$, there is a degeneracy between $Y_{22}$ and a ``cyclic state". Dipole energy will orient the angular momentum axis.  The gap function can be determined by the angular dependence of the momentum distribution of the fermions. 
\end{abstract}
\maketitle

The discovery of fermion superfluids near Feshbach resonances is a major development in cold atoms physics~\cite{dis}. Not only the long sought goal of achieving a fermion superfluid has been realized, the resulting condensate exhibits many interesting ``universal" properties
due to the interplay between unitarity scattering and Fermi statistics~\cite{universal}.  
Even though the studies of these superfluids have just begun, 
another new and exciting direction has already been spun off. 
This is the physics of resonances with non-zero {\em orbital} angular momentum. 

While most experiments on Feshbach resonance focus on $s$-wave scattering, there are 
many Feshbach resonances with non-zero momentum accessible by the same energy tuning method using external magnetic fields. 
Recently, Salomon's group at ENS has reported evidence of reversible production of $p$-wave molecules by sweeping a Fermi gas of $^{6}$Li through a $p$-resonance~\cite{p-Salomon}.  It is therefore conceivable that $p$-wave or even higher angular momentum fermion superfluids can be realized in the future. 
Typically, the width of the resonance decreases with increasing angular momentum. One therefore expects that $\ell \neq 0$ superfluids will be harder to observe than their $s$-wave counterpart.  
Whether this can be achieved within current technology remains to be seen. The fact that the width of 
$p$-resonances can be resolved in recent experiments~\cite{p-Salomon} is very encouraging. 
In any case, the difficulty is not an intrinsic one.  We hope that the novel properties of the $\ell \neq 0$ superfluid pointed out below will motivate searches for these new superfluids. 

In this paper, we consider $\ell \neq 0$   pairing near a  scattering resonance where interaction between particles is strongest.  Our goal is to determine the ground state structure, their signature, 
and the possible existence of universal behavior in these systems.  As a first step, we shall focus at $T=0$. 
In the case of $s$-wave resonance, it has been shown~\cite{Randeria} that mean field theory is valid at $T=0$, despite large fluctuation effects near $T_{c}$. 
Using a model potential that reproduces the exact two-body scattering amplitude in vacuum, we have 
found analytic solutions of the BCS problem for all $\ell \neq 0$ pairing at $T=0$.  Our findings are: {\bf (A)} For $p$ and $f$-wave pairing, the pairing states are $Y_{11}$ and $Y_{32}$ respectively, which are 
orbital ferromagnets that break time reversal symmetry and carry macroscopic angular momenta. 
For $d$-wave  pairing, there is a degeneracy between $Y_{22}$ and a so-called  ``cyclic" state. 
{\bf (B)} The criterion for determining the ground state structure of a $\ell \neq 0$ superfluid is that the energy gap must have minimum angular fluctuation. 
{\bf (C)} Unlike $s$-wave superfluids, where energy per particle has a universal form  
${\cal E}^{}_{F}(1+ \beta)$, the properties of $\ell \neq 0$ superfluids are not universal and are determined by the effective range $r_{\ell}^{}$ of two body scattering.  
The energy per particle is proportional to ${\cal E}_{F}|k_{F}r_{\ell}|<<{\cal E}_{F}$, where 
$k^{}_{F}$ is the Fermi wavevector. 
{\bf (D)} Although dipolar energy is insignificant for superfuid pairing, it breaks the rotational symmetry and orients the angular momentum of the pair.  {\bf (E)}  Experimentally, the nature of the pairing state can be easily revealed by the angular dependence of the momentum distribution of the fermions, whose orientation can be controlled by an external magnetic field through the dipole interaction. 
 These results are established below. 

To begin, we first point out a major difference between the $\ell\neq 0 $ pairing of atomic gases and the $p$-wave pairing of superfluid $^{3}$He.
In the latter case, the pairing interaction is rotationally invariant in spin space.  The $p$-wave interactions between $^{3}$He pairs $(\uparrow\uparrow), (\uparrow\downarrow + \downarrow\uparrow), (\downarrow\downarrow)$ are identical. In the recent ENS experiment, three different $p$-resonances are found for the $^{6}$Li pairs in spin states $|\frac{1}{2}\frac{1}{2}\rangle$, 
 $|\frac{1}{2}\frac{-1}{2}\rangle$, and $|\frac{-1}{2}\frac{-1}{2}\rangle$.  When the pair
 $|\frac{1}{2}\frac{-1}{2}\rangle$ is at resonance, the pairs $|\frac{1}{2}\frac{1}{2}\rangle$, 
and $|\frac{-1}{2}\frac{-1}{2}\rangle$ are not and their interactions can be ignored. 
In other words, particle interaction is highly anisotropic in the ``pseudo-spin" space for atomic Fermi gases near resonance.  

{\bf Setting up the pairing problem: }
We consider a two component Fermi gas (denoted as 
$\uparrow$ and $\downarrow$) with a Hamiltonian $\hat{H}=\hat{H}_{o}+\hat{V}$,
 $\hat{H}_{o}=  \sum^{}_{{\bf k}, \sigma= \uparrow, \downarrow} \epsilon^{}_{\bf k} a^{\dagger}_{{\bf k}, \sigma}a^{}_{{\bf k}, \sigma}$,  $\epsilon^{}_{\bf k} = \hbar^2 k^2/2M$, $M$ is the mass of the fermion, 
$\hat{V}= \Omega^{-1}\sum_{\bf q}\sum_{\bf k, k'}V^{}_{\bf k'-k}a^{\dagger}_{{\bf q/2+k'}\uparrow}
a^{\dagger}_{{\bf q/2-k'}\downarrow} a^{}_{{\bf q/2-k}\downarrow}a^{}_{{\bf q/2+k}\uparrow}$,
where $V_{\bf q}=\int {\rm d}{\bf r} e^{-i{\bf q}\cdot {\bf r}}V(r)$ is the Fourier transform of  the potential $V(r)$ between unlike fermions, and $\Omega$ is the volume. Interactions between like fermions will be set to zero.  If $b$ is the range of the potential $V(r)$, then for wavevector $k$ such that $kb<<1$, the two body scattering amplitude in the $\ell$-th partial wave channel is~\cite{LL}
\begin{equation}
f^{}_{\ell}(k) = \frac{(kb)^{2\ell}}{-a^{-1}_{\ell} + r_{\ell}k^2/2 - i(kb)^{2\ell} k}, 
\label{fellk} \end{equation}
where $a^{}_{\ell}$ and $r^{}_{\ell}$  are the scattering length and effective range respectively. 
At resonance, $a_{\ell}$ diverges, while $r_{\ell}$ remains of microscopic size.  In the case of a square well ($V(r) = -|V|$ for $r<b$, and 0 otherwise), it is straightforward to show that near resonance $r_{\ell}^{} = -b \frac{\ell +1/2}{\ell - 1/2}(2\ell -1)!!^{2}$. 
Recall also that the bound state energy $E_{b}^{}$ will appear as a pole in $f(k)$ when $k$ is continued analytically to the pure imaginary axis, $k \rightarrow i\kappa$. Near resonance, we have 
\begin{equation}
E_{b}^{} = {2\hbar^2\over  M a_{\ell}r_{\ell}}. 
\label{bound} \end{equation}
It is clear that a bound state exists only when $a_{\ell}^{} r^{}_{\ell}<0$.

The scattering amplitude $f_{\ell}(k)$ is related to $V(r)$ through the $T$-matrix $T_{\ell}(k,k';E)$, 
 \begin{equation}
 f_{\ell}(k) =  -\frac{M}{4\pi \hbar^2} T_{\ell}(k, k;  2\epsilon_{\bf k}^{}+ i0^{+}),
\label{fellTell} \end{equation}
where  $T_{\ell}(k,k';E)$ satisfies the integral equation
\begin{equation}
T_{\ell}(k,k'; E_{+}) =  V_{\ell}(k, k') +  \int \frac{{\rm d}p}{2\pi^2} \frac{ p^{2} V_{\ell}(k, p) 
T_{\ell}(p, k'; E_{+}) }{E_{+}  - \hbar^2 p^{2}/M},
 \label{T} \end{equation}
with $E_{+} = $ $E$ $+$ $ i0^{+}$, where we have used the expansion
$V_{\bf k' -k} $ $= 4\pi\sum_{\ell}$ $ V_{\ell}(k,k')$ $\sum_{m=-\ell}^{\ell}$ $Y^{}_{\ell m}(\hat{\bf k})  $ 
$Y^{\ast}_{\ell m}(\hat{\bf k'})$. 

In standard BCS theory,  the ground states is $|G\rangle = \prod_{\bf k} (u_{\bf k} + v_{\bf k} a^{\dagger}_{\bf k, \uparrow}  a^{\dagger}_{-\bf k, \downarrow})|0\rangle $,  
$|u_{\bf k}|^2 + |v_{\bf k}|^2 =1$. The coherence factor $u_{\bf k} $
and $v_{\bf k}$ are determined by minimizing 
\begin{equation}
\langle H-\mu N\rangle = 2 \sum_{\bf k }(\epsilon_{\bf k} -\mu)|v_{\bf k}|^2 - 
\Omega^{-1} \sum_{{\bf k}, {\bf k'}} V_{\bf k' - k}u^{}_{\bf k'}v^{\ast}_{\bf k'} u^{\ast}_{\bf k}v^{}_{\bf k},
\label{grand} \end{equation}
which gives
$u_{\bf k}^{} = \sqrt{ \left( 1 +  \xi_{\bf k}^{}/E^{}_{\bf k}\right)/2 }$, 
$v_{\bf k}^{}/u_{\bf k}^{}= \Delta^{}_{\bf k}/ (E^{}_{\bf k} + \xi^{}_{\bf k})$, 
$E_{\bf k} = \sqrt{  \xi_{\bf k}^2 + |\Delta_{\bf k}|^2}$, 
$\xi^{}_{\bf k} \equiv  \epsilon_{\bf k}^{} -\mu$, 
and the energy gap $\Delta_{\bf k}^{}$ satisfies
 \begin{equation}
\Delta_{\bf k} = -  \sum_{\bf p} V_{\bf k-p}u_{\bf k'} v_{\bf p}  = 
-  \sum_{\bf p} \frac{V_{\bf k-p} \Delta_{\bf p}}{2E_{\bf p}}. 
\label{gap} \end{equation}
The chemical potential $\mu$ is determined by the number constraint
 \begin{equation}
n = \frac{N}{\Omega} =  \frac{1}{\Omega}\sum_{\bf k} \left(  1 - \frac{\epsilon_{\bf k} - \mu}{E_{\bf k}} \right). 
\label{number} \end{equation}
Since we are interested in the resonance physics of  the $\ell$-th partial  wave, 
we replace $V_{\bf k' -k}$ simply by its $\ell$-th angular momentum component, i.e. 
$V_{\bf k' -k} \rightarrow 4\pi V_{\ell}(k,k')\sum_{m=-\ell}^{\ell}Y^{}_{\ell m}(\hat{\bf k})
 Y^{\ast}_{\ell m}(\hat{\bf k'})$.

Even with this (single harmonic)  simplification for $V_{\bf k-k'}$,  there is a serious complication which does not occur  in $s$-wave pairing $--$ that eq.(\ref{gap}) does not have a single harmonic solution, since nonlinearity will force $\Delta_{\bf k}$ to have all spherical harmonics. 
While one expects on physical grounds that only a few harmonics around $\ell$ are dominant, there is no simple way to extract such dominant piece and to calculate the dominant energy contribution.  

To eliminate this technical complication, we adopt the viewpoint that all microscopic potentials that produce  the same low energy scattering amplitude will describe the same low energy 
physics for the system. We can therefore replace the actual $V_{\bf k, k'}$ by a model potential that produces the same scattering amplitude $f^{}_{\ell}$, but allows a much easier solution for the gap equation.  The most convenient model is a generalization of the separable potential
used by Nozieres and S. Schmitt-Rink~\cite{NS}, 
\begin{equation}
V_{\ell}(k,k') = \lambda_{\ell} w^{}_{\ell}(k) w^{}_{\ell}(k'), 
\label{Vseparable} \end{equation}
\begin{equation}
w_{\ell}(k) = \frac{ (k/k_{o})^{\ell}}{[1 + (k/k_{o})^{2}]^{(\ell + 1)/2} }, 
\label{w} \end{equation}
where $k_{o}$ is a momentum cutoff.
With eq.(\ref{Vseparable}),   eq.(\ref{T}) has the solution
\begin{equation}
T_{\ell}^{}(k, k'; E_{+}) = t_{\ell}(E_{+})  w_{\ell}(k) w_{\ell}(k'), 
\label{Tseparable} \end{equation}
\begin{equation}
\frac{1}{t_{\ell}(E_{+})} = \frac{1}{\lambda_{\ell}} - \frac{1}{\Omega} \sum_{\bf k}  \frac{ w_{\ell}(k)^2 }{ E_{+} - 2\epsilon_{\bf k}}. 
\label{t-1} \end{equation}  
For $\ell \neq 0$, eq.(\ref{t-1}) has a low energy expansion~\cite{comment1}, 
\begin{eqnarray}
\frac{1}{t_{\ell}(E_{+})}  & = & \frac{1}{\lambda_{\ell}} +  \frac{1}{\Omega} \sum_{\bf k}  \frac{ w_{\ell}(k)^2 }{2\epsilon_{\bf k}} 
 +  \frac{E}{\Omega} \sum_{\bf k}  \frac{ w_{\ell}(k)^2 }{ 4\epsilon_{\bf k}^2} + O(E^2..) \nonumber \\   &  & 
  + \frac{i\pi}{\Omega}\sum_{\bf k} w_{\ell}(k)^2 \delta(E-2\epsilon_{\bf k}). 
\label{newt-1} \end{eqnarray}  

Substituting eq.(\ref{Tseparable}) and (\ref{newt-1}) into eq.(\ref{fellTell}), and noting that the last term in eq.(\ref{newt-1}) integrates to $ik w_{\ell}^{}(k)^2 (4\pi \hbar^2/M)^{-1}$,  
we achieve the form eq.(\ref{fellk}) provided $\lambda$ and $k_{o}$ are related to the physical 
parameters $a_{\ell}^{}$ and $r_{\ell}^{}$ as 
\begin{equation}
\frac{M}{4\pi \hbar^2 a_{\ell}(k_0 b)^{2\ell}}  = \frac{1}{\lambda_{\ell}}  + \frac{1}{\Omega}\sum_{\bf p} 
\frac{ w_{\ell}(p)^2 }{2\epsilon_{\bf p}} \equiv \frac{1}{g_{\ell}},  
\label{g-1} \end{equation}
\begin{equation}
 r_{\ell} =  -  \frac{2\pi (k_0 b)^{2\ell}}{\Omega} \sum_{\bf p} 
\frac{ w_{\ell}(p)^2 }{\epsilon_{\bf p}^2} \left( \frac{\hbar^2}{M}\right)^2 - \frac{ 2(\ell +1)}{a_{\ell} k^{2}_{o}} 
\label{rell} \end{equation}
More explicitly, we have 
\begin{equation}
 \frac{1}{\lambda_{\ell}}+  \frac{J^{}}{2\pi^2} \frac{M k^{}_{o}}{\hbar^2} = \frac{1}{g_{\ell}}, \,\,\,\,
  r_{\ell} = - \frac{1}{ k^{}_{o}} \left[  \frac{4 I_{1}^{}(k_0 b)^{2\ell}}{\pi}  - \frac{ 2(\ell +1)}{a_{\ell} k^{}_{o}} 
 \right]
\label{newrelation} \end{equation}  
\begin{equation}
J = \int^{\infty}_{0} \frac{ q^{2\ell} {\rm d}q  }{ (1+ q^2)^{\ell + 1} }, \,\,\,\,\,\,
     I_{n}^{} = \int_{0}^{\infty} \frac{ q^{(2\ell-2) n}  {\rm d}q}{ (1+ q^{2})^{(\ell +1)n} }. 
\label{In} \end{equation}  
Near resonance, $a_{\ell}^{}\rightarrow \infty$, the second term in $r^{}_{\ell}$ in eq.(\ref{rell}) and (\ref{newrelation}) can be ignored. 

{\bf Solution of gap equation: }  With the potential given by (\ref{Vseparable}),  
eq.(\ref{gap}) has a solution 
\begin{equation}
\Delta_{\bf k} = w^{}_{\ell}(k) \sum_{m} C_{m} Y_{\ell m}(\hat{k}),
\label{Delta1}\end{equation}
where the coefficients $\{ C_{m} \}$ satisfy the equations
$- C_{m}/\lambda_{\ell}$ $= $ $4\pi\Omega^{-1}$ $\sum_{{\bf p}, m'}   w^{}_{\ell}(p)^2$ $Y_{\ell m}^{\ast}(\hat{\bf p})$ $Y_{\ell m'}(\hat{\bf p})$ $C_{m'}/$ $(2E_{\bf p})$.
Using eq.(\ref{g-1}) to express $\lambda_{\ell}$ in terms of physical parameters $a_{\ell}^{}$ and $r^{}_{\ell}$, we recast this equation as  
\begin{equation}
- \frac{C_{m}}{g^{}_{\ell}}=  \frac{4\pi}{\Omega} \sum_{{\bf p}, m'}  w^{}_{\ell}(p)^2   Y_{\ell m}^{\ast}(\hat{\bf p}) Y_{\ell m'}(\hat{\bf p}) C_{m'}  \left( \frac{1}{2E_{\bf p}} - \frac{1}{2\epsilon_{\bf p}} \right). 
\label{newgap} \end{equation}
The energy of the system, from eq.(\ref{grand}), is
\begin{equation}
\frac{\langle H \rangle}{\Omega} = \frac{1}{\Omega}\sum_{\bf k} \left[ \xi_{\bf k}^{} - E_{\bf k}^{}
+ \frac{ |\Delta_{\bf k}|^2}{2E_{\bf k}^{}} \right] + \mu n. 
\label{energy} \end{equation}

It is useful to separate the angular structure and magnitude of $\Delta_{\bf k}^{}$ by writing
\begin{equation}
\Delta_{\bf k} = w^{}_{\ell}(k) C \tilde{\Delta}(\hat{\bf k}), \,\,\,\,\,\,
 \tilde{\Delta}(\hat{\bf k}) = \sum_{m} \alpha_{m} Y_{\ell m}(\hat{\bf k}) , 
\end{equation}
where $C^2 = \sum_{m} |C_{m}|^2$ and $\alpha_{m}= C_{m}/C$. 
Eq.(\ref{newgap}) can then be written as  
\begin{equation}
- \frac{ C^2 }{4 \pi g^{}_{\ell}} =  \frac{1}{\Omega} \sum_{\bf p} 
|\Delta_{\bf p}^{}|^2  \left(  \frac{1}{2E_{\bf p}} - \frac{1}{2\epsilon_{\bf p}}
 \right).
\label{newnewgap} \end{equation}
Our goal is to solve  from eq.(\ref{newnewgap}) and (\ref{number}) the quantities  ($C$ and $\tilde{\Delta}(\hat{\bf k})$ and $\mu$) as a function of $(n, g_{\ell}^{}$, and
$r_{\ell}^{})$, and then determine which solution
($C$, $\tilde{\Delta}(\hat{\bf k})$) minimizes eq.(\ref{energy}). 

In the following, we shall present an analytic solution for eq.(\ref{newnewgap}) and (\ref{number}) near resonance. We have also solved these equations numerically for all $g_{\ell}$. 
For all the angular momenta we have studied, the two results are {\em identical within the region of $g_{\ell}$ where the analytic solution is valid}.   To derive the analytic solution, 
we write  eq.(\ref{newnewgap}) and  (\ref{number}) in dimensionless form by expressing all energies and wave-vectors in units of 
$\epsilon_{k_{o}}$ and $k_{o}$, i.e.  defining  $\mu \equiv  \overline{\mu} \epsilon_{k_{o}}$, $C \equiv \overline{C} \epsilon_{k_{o}}$,  
$k \equiv  q k_{o}$,  $\epsilon_{\bf k} \equiv   q^2 \epsilon_{k_{o}}$, 
$\Delta_{\bf k}^{} = \epsilon_{k_{o}} w_{\ell}(q)\overline{C}$ 
$\tilde{\Delta}(\hat{\bf q})$. 
Next, we assume that the solution  of eq.(\ref{newnewgap}) and  (\ref{number}) for $\ell \neq 0$  
satisfy  
$\overline{\mu},  \overline{C}^2 <<1$. We will verify later that this is indeed the case when $k_{F}^{}r^{}_{\ell}$  $|r_{\ell}/a_{\ell}^{}| << 1$, 
With this assumption, we can expand eq.(\ref{newnewgap}) and (\ref{number}) in $\overline{\mu}$ and $\overline{C}^2$. 
To the lowest order in these quantity, eq.(\ref{newnewgap}), (\ref{number}) and  (\ref{energy}) 
(when expressed in terms of un-scaled variables) become
\begin{equation}
- \frac{|C|^2}{4\pi g^{}_{\ell}} =  \frac{1}{\Omega} \sum_{\bf p} \frac{ |\Delta_{\bf p}|^2}{2 \epsilon_{\bf  p}^2} \mu  
-  \frac{1}{\Omega} \sum_{\bf p} \frac{ |\Delta_{\bf p}|^4}{4 \epsilon_{\bf p}^3} , 
\label{expandgap} \end{equation}
\begin{equation}
n =  \frac{1}{\Omega} \sum_{\bf p} \frac{ |\Delta_{\bf p}|^2}{2 \epsilon_{\bf  p}^2}. 
\label{expandnumber} \end{equation}
\begin{equation}
\frac{\langle H \rangle}{\Omega} = \mu n  -  \frac{1}{\Omega}\sum_{\bf p}\frac{|\Delta_{\bf p}|^4}{8\epsilon_{\bf p}^3}. 
\label{newenergy} \end{equation}
In deriving eq.(\ref{newenergy}), we need to include second order terms $\overline{\mu}^2$, 
$\overline{\mu} \overline{C}^2$, $\overline{C}^4$ due to cancelation of lower order terms. 
Note that these expansions will not work for $s$-wave since the sums in eq.(\ref{expandgap}) to (\ref{expandnumber}) are infrared divergent. 

Combining eq.(\ref{expandgap}) and (\ref{expandnumber}), we have 
\begin{equation}
n\mu = n\mu_{o} -  \frac{|C|^2}{4\pi g^{}_{\ell}},  \,\,\,\,\,\, 
n \mu_{o} \equiv   \frac{1}{\Omega} \sum_{\bf p} \frac{ |\Delta_{\bf p}|^4}{4 \epsilon_{\bf p}^3} .
\label{nmu} \end{equation}
Using eq.(\ref{expandnumber}) and eq.(\ref{rell}) near resonance,  (and noting that  the $1/(k_{o}a_{\ell})$ term in eq.(\ref{rell}) can be ignored near resonance), we find two equivalent forms for $C$, 
\begin{equation}
 \frac{C^2}{4\pi g^{}_{\ell}} =  \frac{-n}{a_{\ell} r_{\ell}} \frac{\hbar^2}{M}
= -\frac{nE_{b}}{2}, \,\,\,\,\, C = \frac{8(k_0 b)^{\ell}}{\sqrt{3}}\frac{{\cal E}_{F} }{\sqrt{k_{F} |r_{\ell}^{}|}}. 
\label{finalC}  \end{equation}
Comparing eq.(\ref{nmu}) and (\ref{finalC}), we have  
\begin{equation}
\mu = \mu_{o}^{} + E_{b}^{}/2. 
\label{finalmu} \end{equation}
 The explicit form of $\mu^{}_{o}$ can be obtained by  evaluating the integral in eq.(\ref{nmu}) 
 and using the expression of $r_{\ell}^{}$ in eq.(\ref{newrelation}) near resonance, which is 
 \begin{equation}
\mu_{o} =\gamma  \left(  \frac{2\pi^2}{3(k_0 b)^{2\ell}} \frac{ I_{2}}{I_{1}^{3}}  
\right)  {\cal E}_{F}^{} |k_{F}^{} r_{\ell}|, \,\,\,\,\,\, \gamma = \int {\rm d}\hat{\bf p}|\tilde{\Delta}(\hat{\bf p})|^4
\label{muo} \end{equation}
where $I_{n}$'s are given in eq.(\ref{In}). 
Eq.(\ref{finalC})  to  (\ref{muo}) together with  eq.(\ref{bound}) give $C$ and $\mu$ as a function of $n$, $a_{\ell}^{}$ 
and $r_{\ell}^{}$.  From these equations, it is easy to show that  $\overline{\mu}, \overline{C}^2\sim |k^{}_{F}r^{}_{\ell}|^3$. Since  $|k_{F}^{}r_{\ell}|<<1$,  our initial assumption  $\overline{\mu}, \overline{C}^2<<1$ is valid.  Note that the structure of the gap $\gamma$ shows up only in $\mu$ and not in $C$. 
 
Finally, using eq.(\ref{newenergy}), (\ref{nmu}), and (\ref{finalmu}), 
we obtain the energy density as a function of $n$, $a_{\ell}^{}$ and $r_{\ell}^{}$, 
\begin{equation}
\langle  H \rangle/\Omega = n \mu - n\mu_{o}/2 = (n/2)(\mu_{o} + E_{b}).
\label{finalenergy} \end{equation}
Eq.(\ref{finalenergy}), (\ref{muo}), and (\ref{bound}) imply that 
(i) the ground state has minimum angular fluctuation in the gap, i.e. $\gamma$.  (ii) Unlike the $s$-wave case where $\mu$ is of order ${\cal E}_{F}^{}$ near resonance,  eq.(\ref{muo}) shows that the chemical potential $\mu$ 
 for $\ell\neq 0$ at resonance is greatly reduced from ${\cal E}_{F}$,  by a {\em non-universal} factor $|k_{F}^{}r_{\ell}|$, reflecting a stronger interaction energy then the s-wave case. This is due to the fact that the energy of the bound state $E_{b}$ for $\ell \neq 0$ grows much faster than that of $s$-wave away from resonance, since $E_{b}\propto 1/(a_{\ell}^{} r_{\ell}^{})$ for $\ell \neq 0$ whereas $E_{b}\propto -1/a_{0}^2$ for $\ell = 0$.  The effect of this larger interaction energy also shows up at high temperatures. However, the thermodynamics in that regime is universal~\cite{HZ}.  

\begin{figure}[t]
\centering \epsfig{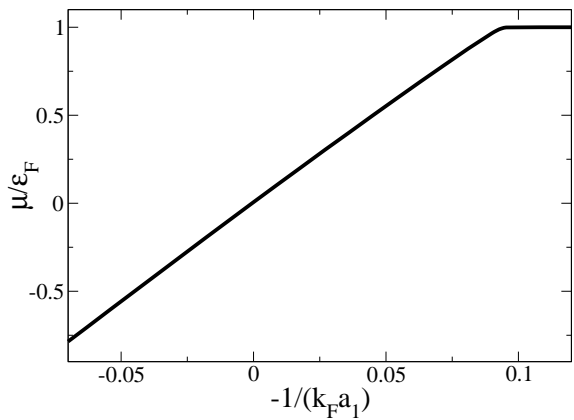}
\caption{Chemical potential ($\mu$) for $\ell=1$ resonance as a function of $-1/k_{F}a_{1}$. 
We have taken $r_{1} = - 3b$, $k_{o}$ given by eq.(\ref{newrelation}), and $k_{F}r^{}_{1}= -0.09$. The linear portion agrees exactly with eq.(\ref{finalmu}) and (\ref{muo}). The switching to $\mu = {\cal E}_{F}$ at the atomic side is very sudden and is controlled by the smallness of $k_{F}^{}r^{}_{1}$. It takes place around $E_{b}/2 \sim {\cal E}_{F}^{}$, or $(k^{}_{F}a^{}_{1})^{-1} \sim k_{F}r^{}_{1}$.} \label{figure1.fig}
\end{figure}

{\bf  Ground state structure:}  To find the minimum of $\gamma$, we shall use the rectangular representation of spherical harmonics to write $\sum_{m} \alpha_{m} Y_{\ell m}(\hat{\bf k})$ $= \sum_{[i]} A_{i_{1}, i_{2}, ..., i_{\ell}}k_{i_{1}}^{}
k_{i_{2}}^{}... k_{i_{\ell}}^{}$, where $A$ is symmetric in all its indices, and vanishes upon contraction of any two pairs of indices.   For example, the sum $ \sum_{m} \alpha_{m} Y_{\ell m}(\hat{\bf k})$ is 
${\bf A}\cdot \hat{\bf k}$, $A_{a,b}\hat{k}_{a}\hat{k}_{b}$, $A_{a,b,c}\hat{k}_{a}\hat{k}_{b}\hat{k}_{c}$, for 
$\ell=1, 2$, and $3$ respectively, with ${\rm Tr}A=0$ for $\ell=2$ and $\sum_{a}A_{a,a,c}=0$ for $\ell=3$.  We then have 
$\gamma_{_{\ell =1}}^{} \propto 2 ({\bf A}^{\ast}\cdot {\bf A})^2 + | {\bf A}^{}\cdot {\bf A}|^2$,
$\gamma_{_{\ell =2}}^{}   \propto  2 ({\rm Tr} A^{\dagger}A)^2 + |{\rm Tr} A^2|^2$, 
and a similar but lengthier expression for $\ell=3$. 

For $\ell=1$,  the minimum occurs at ${\bf A}^2=0$.  This means ${\bf A} \propto \hat{\bf x} + i\hat{\bf y}$ or any rotation of it, which is  $\tilde{\Delta}(\hat{\bf k}) =Y_{11}(\hat{\bf k})$ along an arbitrary angular momentum quantization axis. 
The superfluid is an ``orbital ferromagnet"  since all pairs are in the same $Y_{11}$ state.  This is remarkable for it implies that by sweeping across the Feshbach resonance, a superfluid with macroscopic angular momentum and broken time reversal symmetry will result. 

The problems of minimizing $\gamma$ for $\ell=2,3$ were solved by N.D. Mermin  in the context of superfluid $^{3}$He~\cite{Mermind,Merminf}.   For $\ell=2$, there is an accidental degeneracy. Both $\tilde{\Delta}(\hat{\bf k})=Y_{22}(\hat{\bf k})$ and the ``cyclic" state $\tilde{\Delta}(\hat{\bf k})\propto \hat{k}^{2}_{x} + e^{2\pi i/3}  \hat{k}^{2}_{y}
+  e^{4\pi i/3}  \hat{k}^{2}_{z}$ minimize $\gamma_{\ell =2}^{}$~\cite{Mermind}.    This degeneracy can be resolved in higher order in $|k_{F}r_{2}|$ and will be discussed elsewhere.  For $\ell=3$, the ground state is $\tilde{\Delta}(\hat{\bf k}) =Y_{32}(\hat{\bf k})$ along an arbitrary direction~\cite{Merminf}. This state is also an orbital ferromagnet, even though it is not of maximum angular momentum state.  At present, there are no exact solutions for $\ell \geq 4$. 

{\bf Numerical Results:} Note that although $\mu\sim {\cal E}^{}_{F}|k_{F}^{}r_{\ell}^{}|$ near resonance (eq.(\ref{muo})), 
it has to recover to $\mu \sim {\cal E}^{}_{F}$ on the atomic side of the resonance whether $a_{\ell}^{}$ is negative and small. We have solved eq.(\ref{newnewgap}) and (\ref{number}) numerically and have shown this is the case. (See figure 1). Our numerical results are in exact agreement with eq.(\ref{finalmu}) and (\ref{muo}) near resonance.

{\bf The effect of dipolar energy:} Dipolar energy $V_{D}^{}$ breaks rotational symmetry in real space. 
Since electron spins are polarized by the external magnetic field ${\bf B}$, we have 
$V_{D} =\frac{1}{2}\int U({\bf r}- {\bf r'}) \psi^{\dagger}_{\alpha}({\bf r}) 
\psi^{\dagger}_{\beta}({\bf r'}) \psi^{}_{\beta}({\bf r'}) \psi^{}_{\alpha}({\bf r})$, 
$U({\bf r})  = \mu_{B}^2(1-3(\hat{\bf B}\cdot \hat{\bf r})^2)/r^{3}$, where 
 $\mu_{B}$ is the electron Bohr magneton.  Since dipolar energy per particle is $\mu_{B}^2 n$, and since 
$\mu_{B}^2 n/\mu =  (e^2 k_{o}/m_{e}c^2) \left(M/m_{e}\right) (3\pi^2/2) \sim 10^{-4}$
for $k_{o}\sim 10^4 {\rm cm}^{-1}$, dipolar energy is not strong enough to affect the gap structure. 
On the other hand, it can orient the pairing state.  A straightforward calculation shows that 
$\langle V_{D}\rangle = (2\pi\mu_{B}^{2}/\Omega)\sum_{\bf k, k'} ( 1/3 - (\hat{\bf B}\cdot \hat{\bf q})^2)
\Delta_{\bf k'}^{\ast} \Delta_{\bf k}^{}/(4E_{\bf k}^{} E_{\bf k'}^{})$, where ${\bf q} = {\bf k'} - {\bf k}$. 
In the case $\ell=1$, $\Delta_{\bf k}^{} \propto {\bf A} \cdot {\bf k}$, 
we have  $\langle V_{D}\rangle \propto   - |\hat{\bf  B} \cdot {\bf A}|^2$. Since 
${\bf A} \propto \hat{\bf x} + i \hat{\bf y}$, we have  $\langle V_{D}\rangle \propto   +( \hat{\bf  B} \cdot \hat{\bf z})^2$.  The angular momentum of the pair will lie in the plane perpendicular to ${\bf B}$.

{\bf Signature of the $\ell\neq 0$ superfluid:}   Eq.(\ref{expandnumber}) shows that the momentum distribution $n_{\bf p}$ of the fermions is $|\Delta_{\bf p}|^2/(2\epsilon^{2}_{\bf p})\propto|\tilde{\Delta}(\hat{\bf p})|^2$.  A measurement of  the angular dependence of $n_{\bf p}$
therefore gives  $|\tilde{\Delta}(\hat{\bf p})|^2$ directly. 

We have thus established results ${\bf (A)}$ to ${\bf (E)}$ mentioned in the Introduction.  
This work is supported by NASA GRANT-NAG8-1765  and NSF Grant DMR-0109255.



\begin{thebibliography}{99}
\bibitem{dis}  C. A. Regal, M. Greiner, and D. S. Jin, Phys. Rev. Lett. {\bf 92}, 040403 (2004). 
M. Zwierlein, et al., Phys. Rev. Lett. {\bf 92}, 120403 (2004). M. Bartenstein, et al, cond-mat/0403716. 
J. Kinast, et al  Phys. Rev. Lett. 92, 150402 (2004).  C. Chin, et.al. cond-mat/0405632. 
M. Greiner, C. A. Regal,  and D. S. Jin, cond-mat/0407381.
\bibitem{universal} K. M. O'Hara, et al.,  {\em Science} {\bf 298}, 2179 (2002). M.E. Gehm, et al., Phys. Rev. A 68, 011401(R) (2003). T. Bourdel, Phys. Rev. Lett. 91, 020402 (2003). 
\bibitem{p-Salomon} J. Zhang, et.al, quant-ph/0406085
\bibitem{Randeria}  C. S‡ de Melo, M. Randeria, and J. Engelbrecht, 
 Phys. Rev. Lett. {\bf 71}, 3202 (1993)
\bibitem{LL}  p.556 in {\em Quantum Mechanics}, 3rd ed., by L.D. Landau and E.M. Lifshitz, 
Butterworth and Heinemann, 2002. 
\bibitem{NS}  P. Nozieres and S. Schmitt-Rink, 
J. Low Temp. Phys. {\bf 59}, 195 (1985). 
\bibitem{comment1} This expansion does not apply to $\ell=0$ because the first two integrals diverge. The integral in eq.(\ref{t-1}), however, can be calculated analytically for all $\ell$. The result is not displayed here because it is not illuminating. The analytic result for $\ell \neq 0$ agrees with 
the expansion eq.(\ref{newt-1}). 
\bibitem{HZ} T.L. Ho and N. Zahariev, cond-mat/2004. 
\bibitem{Mermind} N.D. Mermin, Phys. Rev. {\bf A}, {\bf 9}, 868, (1974). 
\bibitem{Merminf} N.D. Mermin, Phys. Rev. {\bf B}, {\bf 13}, 112, (1976). 
\end{thebibliography}
\end{document}